# SPECTRUM AUCTIONS, PRICING AND NETWORK EXPANSION IN WIRELESS TELECOMMUNICATIONS

Johannes M. Bauer[*]

## 1. Introduction

This paper examines the effects of licensing conditions, in particular of spectrum fees, on the pricing and diffusion of mobile communications services. Seemingly exorbitant sums paid for 3G licenses in the UK, Germany in 2000 and similarly high fees paid by U.S. carriers in the re-auctioning of PCS licenses early in 2001 raised concerns as to the impacts of the market entry regime on the mobile communications market. In a very short time, two rival positions emerged among policy makers and researchers.

On the one hand it is argued that license fees are sunk investment costs and therefore do not influence the subsequent evolution of the wireless market (Cave and Valletti, 2000). Properly designed auctions are seen as the most efficient way of managing spectrum. On the other hand it is argued that governments may lack the information to design efficient auctions. Furthermore, they may be inclined to abuse spectrum auctions to extract rent from the wireless industry. As a result, auctions will tend to reduce the number of competitors in the market, increase the level of prices, and slow down network deployment (Noam 1998, Gruber 2001, Melody, 2001).

Concern about the potential negative impacts of license fees has led to a review of the public policies towards 3G infrastructure deployment in several European nations. For example, in March 2001 the Commission of the European Communities issued a statement encouraging national regulatory authorities (NRAs) to explore possibilities to mitigate the burden of investment for 3G licensees (European Commission 2001). Among the policy options mentioned are the sharing of infrastructure, as well as extended payment schemes or discounts for the license fees. Several national regulatory agencies are also reconsidering their regulatory policy. For example, the Germany regulator (RegTP) is currently reviewing industry proposals for network sharing and delayed license fee payment (Financial Times, April 3, 2001).

[*] Associate Professor, Department of Telecommunication, Michigan State University, 409 Communication Arts and Sciences, East Lansing, MI 48824, USA, email bauerj@msu.edu. Part of the background research for this paper was conducted while the author was visiting professor at the Center for Infrastructure Design and Management, Faculty for Technology, Policy, and Management, Delft University of Technology.

The current regulatory debate seems driven by instincts rather than serious analysis. To a certain degree, re-regulation seems to be a reaction to earlier regulatory failure in the design and coordination of national licensing processes. It also seems to be a reaction to the surge in debt taken on by mobile service providers and the corresponding decline in stock values. It must be doubted whether these are proper instances for regulatory intervention. Moreover, the discussed measures have risks of their own. Some of the proposed relief measures, such as infrastructure sharing, may reduce the effectiveness of competition in future 3G markets. Other options such as discounts on the winning bids, may reduce the future credibility of spectrum auctions and thus cause long run damage.

This paper intends to contribute to a clarification of these issues in two ways. First, the paper models the nexus between the market entry regime and the subsequent pricing and investment decisions of a licensee. We analyze the possible implications of the license regime under alternative behavioral assumptions for the licensees. Taking advantage of the fact that licensing regimes vary across countries, we are able to test our model using data for GSM and PCS networks and services. The data provides qualified support for the hypothesis that license fees may increase the prices of wireless services and slow down the growth of the industry. Based on this analysis, the paper, second, explores the implications of these findings for the current 3G debate.

The next section of the paper briefly reviews licensing regimes for mobile services. Section three develops a stylized model of the potential implications of license fees on the pricing of services and the growth of the market. It also derives two testable hypotheses. Section four discusses the empirical model utilized to test key hypotheses as well as key findings. Section five analyzes the current 3G regulatory proposals and discusses future policy options and section six summarizes the main insights.

**2. Approaches to spectrum management**

In many ways, electromagnetic spectrum is a unique resource. Unlike other resources, it cannot be depleted. While the laws of physics limit the range of usable frequencies, its capacity and throughput can be expanded by improved with more efficient technology. However, as electromagnetic waves interfere with each other, strong interdependencies exist among users and uses. As was recognized after an initial chaotic period in the early parts of the 20$^{th}$ century, the orderly use of spectrum requires some form of coordination. While the views as to the most efficient approach differed widely almost from the beginning, national and international administrative procedures were established to allow a more orderly utilization.

The most widespread approach to spectrum management uses a two-step approach. In a first stage, frequency bands ("blocks") are allotted to specific uses such as voice communications. Recently, regulatory agencies have also begun to experiment with more open allotment, which allows spectrum in a certain band to be used for a pre-specified range of services (e.g., voice, data, paging, and Internet access). Certain limited uses, such as the Citizen Band Radio service or low power radio, are often granted by rule and do not require a specific license. The two stages are not interdependent as allocation decisions have important repercussions for the subsequent evolution of a service (Gruber 2001, Melody 2001).

Until recently, regulatory agencies relied on an administrative process to assign licenses to users. A potential licensee typically had to meet certain legal, technical and business qualifications. In cases of competing applications, licensees were selected in a "beauty contest" based on more or less complex sets of criteria, such as the proposed service, the efficiency of spectrum use, or the prior record of an applicant. While administrative licensing allowed the implementation of certain policy goals through the licensing process, it was plagued with difficulties of identifying the best proposal and long delays. In the 1980s, the US Federal Communications Commission experimented with lotteries to shorten the licensing process. Widespread speculation for windfall gains revealed the randomness of spectrum lotteries. As a remedy, Congress in 1993 authorized the use of auctions to improve the efficiency of spectrum management.

Spectrum auctions need to overcome several difficult challenges (Milgrom, 2001). For example, bidding rules need to be designed to prevent collaboration between participants in the auction. As licenses in adjacent territories may be complementary to each other, bidders should be allowed to bid for licenses simultaneously. Last but not least, the rules need to be designed so as to converge to an efficient outcome. This includes procedural rules for the auction process but also the broader obligations attached to a license, such as targets for infrastructure deployment. To cope with the procedural challenges, the FCC, with experts from the academic community, developed a unique method, simultaneous multiple-round auctions (SMRA). In this approach licenses in one service but in different territories are auctioned off at the same time. Based on activity rules that ascertain efficient bidding, the auction proceeds in rounds. After each round, the new bids are revealed to all participants. The auction ends when no bidder has an incentive to change his bid. Since 1993, the FCC has conducted 38 auctions and collected a total of $41.6 billion (FCC 2001). Many nations have

since adopted auctions as a method of spectrum assignment, especially for the next generation of 3G services.

Nevertheless, there is a great diversity of licensing regimes in place with widely differing impact on the market entry costs of service providers. Tables 1 and 2 summarize key features of the 2G and 3G licensing approaches in selected OECD countries. Whereas 2G services (GSM and PCS) have been introduced in the 1990s and experienced rapid market growth, 3G services are still on the drawing board. Due to difficulties with hardware and software as well as market uncertainty, earlier plans to start commercial operations of 3G services in the spring of 2001 were postponed. Based on current plans, most service providers seem to target 2003 as the introduction date of full-fledged commercial offerings.[1]

[Table 1 about here]

[Table 2 about here]

Tables 1 and 2 show significant differences in the national approaches to the licensing of GSM/PCS and 3G services. Outside of the U.S., GSM licenses were typically awarded in an administrative process. Generally one available license was granted to the incumbent fixed service provider. Significant differences also exist in how countries charge for licenses. Several countries have elected not to charge any or any significant license fees for GSM and even 3G services. Some countries, for example, Denmark, have not charged service providers for GSM licenses but have introduced license fees for 3G. For the award of 3G licenses, more countries have opted to use auctions. The most widespread approach are simultaneous multiple round auctions. Denmark decided to rely on a sealed bid auction in which the price of the four licenses was determined by the fourth bid.[2] Nevertheless, several countries including France and Finland continued to use beauty contests. The magnitude of the differences becomes visible in the per subscriber impact of one time and recurring license fees. In the GSM/PCS service the per subscriber fee ranges from 0 in Germany to $466 in the U.S. (Sprint PCS). In the 3G service the average per subscriber fee is much higher than in the existing services. Fees range from zero in Finland to almost $1,948 per subscriber in Germany.

---

[1] In April 2001, after several months of an international advertising campaign, NTT DoCoMo announced that it would push back the launch of its 3G service from May to October 2001. Other service providers, including Telefónica Moviles or BT, have also delayed the introduction of 3G services.

For the purposes of this paper it is of particular interest whether and how the differences in licensing fees affect post-entry conduct. There are differences with respect to the mix of fixed one-time fees and the reliance on recurring variable payments. In the GSM/PCS service, in most countries licensing fees consist of an administrative and a frequency use component. The administrative fee is typically intended to recover spectrum management costs of the regulatory authority but may also include a one-time license fee for access to spectrum. In contrast, frequency fees are typically assessed to reflect the scarcity of the resource (ETO, 1998, p. 55). Within the U.S. and EU a great variety of payment models and fee structures exist both in the 2G and 3G services. This variance might be used in exploring the impact of market entry fees on the subsequent evolution of the market.

3.  **A conceptual framework**

From an economic point of view, the fees assessed from service providers fall into two broad categories, namely one-time or annual fixed payments and variable payments. Whereas there is agreement that variable cost influence the pricing and investment decisions of firms, there are two competing positions with respect to the impact of one-time license fees. One position emphasizes the sunken nature of such one-time costs. In a competitive market, it is argued, sunk costs will not influence the marginal costs of the service provider and thus not influence the market price. In this view, spectrum auctions are seen as a two-step process. In the first stage, based on conjectures as to the future evolution of the market, firms determine their maximum willingness to pay for a license. Those firms who succeed will base subsequent decisions only on incremental costs and revenues (Cave and Valletti 2000). Thus, the evolution of the market in the second stage is independent of the first stage. There are several potential problems with this argument that we will explore in this section.

---

[2] Anne Young, Denmark awards four 3G licenses, Total Telecom, 20 September 2001.

These weaknesses all have to do with the (implicit) assumption that the two stages of the evolution of the market can be separated. Mobile service providers, like other firms, cannot afford to lose money on an ongoing basis. Therefore, the average cost of producing a service will determine the long-run minimal price at which a firm is willing to offer service. Only in the short run will they be willing to offer services at a marginal cost price below average costs. The average cost of service provision will include the capitalized costs of acquiring a license. This is represented in figure 1. License fees increase the average cost of supplying the service at the firm level and increase prices at the market level by reducing the aggregate supply. Compared to a fully competitive market without entry fees, the new equilibrium is characterized by higher prices and lower quantities sold. If network externalities exist, these higher prices may exert a negative feedback effect and further slow down the overall expansion of the market. As all licensees face similar entry fees such higher prices can be effected through parallel behavior not necessitating any form of tacit or open collusion.

[Figure 1 about here]

Mobile service providers have historically used more differentiated pricing models than their wireline counterparts. For example, pricing structures that offer a block of minutes for a fixed monthly fee combined with a usage charge for excess time are very widespread. Most wireless service providers offer different combinations of fixed and variable charges resulting in lower average prices per minute for customers electing higher volume calling plans.[3] In addition they have some discretion to set roaming charges for calls originating or terminating in other networks. These charges are often rather intransparent to the user at the time of making a call. Last but not least, service providers use other features such as discounts on handsets, minimum contract periods combined with early termination penalty fees. Therefore, mobile service providers have a wide range of options for effecting their prices.

Another intertemporal link between the licensing stage and the development of the market further exists due to the influence of the entry conditions on the structure of a market. Sutton (1991) has provided ample evidence for the fact that higher entry costs contribute to a more highly concentrated industry structure. At first sight, the market structure in wireless services seems exogenously determined by the number of licenses that the government makes available. However, this need not be the equilibrium outcome. Building on Sutton's argument, Gruber (2001) argues that the uncertainty surrounding spectrum auctions may result in bids that are above the competitive equilibrium, intensifying the pressure for consolidation. Moreover, there exist other forms of cooperation that may have a similar effect as open consolidation. In the 3G market, firms have responded to the high license fees with agreements to share network infrastructure. Not only does network sharing run counter the vision of facilities-based competition, it also raises the risk of collusion among the participants.

---

[3] For example, Vodafone in the U.K. offers nine options in the standard service category, ranging from blocks of 20 minutes to 1,000 minutes included in the basic monthly fee. Calls in excess of the allowed minutes are priced at 10-15 pence per minute peak and 5 pence off-peak. A further differentiation is introduced through monthly allowances for text messages. In addition, Vodafone offers a range of leisure plans and special saver plans. See <http://www.vodafone.com/external.htm?url=http://www.vodafone.co.uk> for more details.

The impact of these two effects on prices and network expansion is complicated by the fact that mobile service providers typically have several options for the configuration of their networks. The initial and overall investment costs faced by a firm depend on the coverage and capacity of the network. Coverage and capacity will in turn influence the quality of service and the number of users a network can handle and hence the revenue potential of a supplier. It is not clear a priori how license fees will influence the capacity and coverage choices of firms. However, it is likely that they will result in less coverage and a lower-capacity system, even if later expansion is more expensive than building a higher capacity system from the beginning. If this is the case, the choice of a lower-capacity system design, which will reduce the ratio of the minimal efficient firm size to the overall size of the market, may mitigate the pressure for higher industry concentration.

The effect of different coverage and capacity choices on the average cost of service provision depends on several factors, including the topography of a service territory, demographic factors, and engineering cost factors. A focus on higher density, easier-to-serve territories may reduce the average costs of service provision albeit at the potential cost of constraints on revenue growth. The imposition of license fees will therefore lead to adjustments in the investment behavior of a firm. To avoid ex post adjustments, many regulatory agencies have established network coverage and capacity targets. These counteracting forces determine the net effect of a license fee. Given the constraints on licensees' investment plans it can be expected that license fees will tend to increase the average cost of service.

[Figure 2 about here]

Effects on the pricing of service and network expansion can also result from the fact that not all license-related costs are fixed. For example, high expenses for a license may increase the debt burden of a mobile service provider leading to higher costs of capital for any rate of network expansion. Furthermore, high upfront expenses will reduce cash or investment reserves and thus may tighten the financing constraint on an operator in addition to higher costs of capital. In both cases the result will likely be a slower path of network expansion and/or higher prices for the services provided.

These arguments challenge the conventional view that license fees do not affect the subsequent evolution of the wireless market. In real world wireless markets the impact of license fees on the pricing of services and network expansion is shaped by several other factors. First, second generation wireless (GSM, DCS, PCS) licenses were in most countries issued in a staggered fashion. Typically, one initial license was awarded to the incumbent PTO and one or more to competitors. As the market expanded, new license such as the GSM-1800 licenses in Europe, were awarded. With very few exceptions the early entrants were able to maintain a strong market position (OECD 2001, p. 33). As license fees were typically introduced only for the more recent licenses, late entrants faced the dual disadvantage of a small customer base and higher entry costs. In this context, the effects of license fees may not be visible in higher prices but in lower earnings and valuation of the service providers. Second, rather than result in permanent effects on the pricing of services, in the fast growing wireless markets the effects of license payments could be only temporary (see figure 3). Depending on the speed of adjustment towards such a competitive outcome, the impact of license fees could vary over time. This raises additional empirical issues as service providers may be at different stages in this cycle.

These arguments point to some testable hypotheses but also to some of the challenges that need to be overcome in an empirical assessment of the impacts of license fees. To build an empirical test, we can take advantage of the differences in licensing conditions across nations. The two null hypotheses are that (a) that license fees impact prices in the market for wireless services and (b) that license fees slow the growth of the wireless market. From the conceptual discussion in the previous section follows:

$$\text{If } H_0^a \text{ then } \partial p/\partial C_L > 0 \text{ and } \partial q/\partial C_L < 0 \quad (1)$$

$$\text{If } H_0^b \text{ then } \partial g/\partial C_L < 0 \quad (2)$$

With

$p$ …. price of service

$q$ … quantity

$g$ … growth rate of network

$C_L$ .. license fee

In this paper we test the first hypothesis empirically using a sample of service providers subject to varying licensing regimes.

## 4. Empirical model and findings

In order to isolate the potential influence of license fees demand and supply-side factors impacting the market price as well as network expansion need to be controlled for. It is reasonable to assume that all carriers in a market have access to the same technology, although they may have different skills in integrating this technology in their overall network. Other things equal, it seems reasonable to assume that the quantity supplied will increase with the market price. Moreover, it can be expected that the entire supply curve will be subject to some shift factors. For example, in markets with a higher intensity of competition firms will be forced to reduce X-inefficiency, resulting in a downward shift of the supply curve. In markets with a high population density the supply curve will likely be lower than in markets with sparse population. The demand relation is, in addition to the market price, shaped by the income of consumers, the price of substitutes such as wireline service, and the quality of such services including their availability. For our empirical model we identify the supply and demand relationships as

$$q^S = \alpha_0 + \beta_1 p^W + \beta_2 C_L + \beta_3 COMP + \beta_4 POPD + \beta_5 W + \varepsilon_1 \qquad (3)$$

$$q^D = \alpha_1 + \beta_6 p^W + \beta_7 INC + \beta_8 p^F + \beta_9 TD^F + \varepsilon_2 \qquad (4)$$

In equilibrium the following condition must hold

$$q = q^S = q^D \qquad (5)$$

With

$q^S$ ...    quantity supplied, normalized to correct for differences in the national network sizes

$q^D$ ...    quantity demanded, normalized to correct for differences in the national network sizes

$p^W$ ...    price of wireless voice service, annual cost of representative basket

$p^F$ ...    price for fixed voice service, annual cost of representative basket

$C_L$ ...    license fee per subscriber

COMP ... competition in the market, measured by Herfindahl-Hirschman Index

POPD ... population density in service territory

W ...    measure of national wage level

INC ...    income of consumers, measured as GDP per capita

$TD^F$ ...    teledensity for fixed service

Utilizing (5) after some manipulation we can derive the reduced form of the system as

$$p = \gamma - \xi_1 C_L - \xi_2 COMP - \xi_3 POPD - \xi_4 W + \xi_5 INC + \xi_6 p^F + \xi_7 TD^F + \psi \qquad (6)$$

The structural model represented in equations (3) to (5) can be estimated using three stage least squares (3SLS). Parameter estimates of the reduced form represented in equation (6) can be generated with ordinary least squares (OLS). The first approach yields estimates of the structural parameters of the system whereas the reduced form produces parameter estimates that include both the direct and indirect effects of changes in exogenous variables.

Not all of the exogenous variable can be observed directly and therefore proxies had to be used (see table 3 for summary statistics). The use of a cross-national sample further complicates the issues. Pricing models for telecommunications services vary across countries. Differences exist in the delineation of local and long distance calls, time-of-day pricing, and the availability of discounts. Many of these differences (but not all) can be overcome by using a basket of calls (OECD 1990, 2000). We use the most recent OECD baskets for residential and business mobile and fixed voice services. The basket represents the annual cost of an average user and reflects national and international calls placed at different times of the day as well as discounts (see OECD 2001 for a more detailed description).

[Table 3 about here]

License fees are assessed either as upfront payments, as ongoing usage fees or a combination of both. In a dynamic context, these payment approaches have different implications. Whereas upfront payments increase the minimal efficient scale of operations variable payments decrease it. Thus, it can be expected that upfront payments will increase market concentration whereas variable payments will decrease it. Both types of fees will exert a positive impact on prices. However, this effect may be counteracted by increased competition in a scenario with variable payments. Since we are interested in testing whether license fees do indeed have a positive impact on price, we do not distinguish between these two approaches and calculate the cost of the license as the per subscriber payment over a five year period. Effective competition in the market for wireless services should increase the incentive of service providers to produce and price more efficiently. We measure the intensity of competition using the Herfindahl-Hirschman Index (HHI). To reflect the varying cost conditions in national markets we include a measure of population density and a measure of the national wage level. The former variable is intended to capture the fact that capital costs of building a wireless network are inversely related to population density and positively related to the cost of labor.

The key variables included on the demand-side of the market are the price for wireless services, income, the price of substitute services, and the quality of service of the existing network. We represent income by the GDP per capita in US$ purchasing power parities (PPP). As a proxy for the price of fixed voice service we use the OECD basket for residential and business voice service as of November 2000. Quality of service of the existing voice service is represented by the teledensity of the fixed voice network. The relation between wireless and fixed services is a priori not clear and can be of both a complementary as well as substitutive nature. This relationship may change as the wireless network matures. In some of the Nordic countries consumers have apparently begun to unsubscribe fixed voice service for wireless service. As a result the current specification of the model may not produce meaningful results for the fixed network variables. Table 4 gives an overview of the correlation between exogenous variables.

[Table 4 about here]

The results derived from 3SLS estimation of the structural model for the residential market are summarized in table 5. The equations estimate the quantity supplied or demanded, treating the price for wireless service as endogenous. Due to the small number of observations, the results need to be interpreted with caution. Most coefficients show signs compatible with our theoretical discussion. As expected, on the supply side a higher market price for wireless services corresponds with higher quantities supplied and higher cost of a license result in lower quantities supplied. It is interesting that the competition variable enters with a positive sign, suggesting that a higher degree of competition results in lower quantities offered.[4] This could be interpreted as support for the Schumpeterian notion that there is an optimal degree of market concentration, which facilitates innovation and that more highly fragmented markets are less efficient. However, more work will have to be done to substantiate this conjecture. Population density and wage levels have unexpected signs. This could point to relations between variables that were not explicitly measures. For example, in countries with a lower population density wireless service may be cheaper to implement than fixed wire service thus yielding a negative influence of population density on the quantity supplied. The positive sign of the wage level variable could be explained with a more skilled work force and thus higher labor productivity.

[Table 5 about here]

On the demand side, the price for wireless service has the expected negative sign and income the expected positive sign. The positive sign of the price for fixed services seems to indicate that there is indeed a substitutive relationship between fixed and wireless service. The negative sign of the teledensity variable supports this. It would suggest that, other things equal, wireless service is more widely diffused in countries with a poorer fixed voice system. The specification of the supply relationship explains 62% and the demand relationship 22%

---

[4] Competition is measured by the Herfindahl-Hirschman Index, which ranges between 0 for perfect competition to 10,000 for monopoly. Therefore, a positive sign of the coefficient implies that a more concentrated market

of the variance. Overall, the supply function is significant at the >99% level, whereas the demand function is significant at the >95% level. It would be desirable to increase the number of observations to narrow the confidence intervals of the coefficients.

## 5. Implications for 3G policy

During the past few years, many (but not all) countries have adopted market-based mechanisms to award licenses. Prices offered for 3G spectrum were much higher than the fees paid for GSM and PCS services. This is due to a combination of several factors, including the specific auction designs, a general overheated phase in the evolution of telecommunications markets, and a fairly high degree of uncertainty as to the revenue and profit potential of the 3G market. Moreover, as auctions were conducted at different points in the overall stock market cycle, prices vary strongly between nations. While it will be interesting to conduct an empirical study of the impacts of 3G license payments once data on service pricing and diffusion becomes available, some tentative comments on 3G regulatory approaches can be made based on our analysis.

In Europe, 3G auctions prior to September 2001 grossed more than $130 billion in fees. As 3G is a fully digital technology and requires the installation of a new radio access network. The propagation characteristics and the use of spread spectrum principles in the engineering of 3G networks imply that the topography of an optimally configured network differs from the installed GSM or PCS networks. Thus, new antenna towers will have to be built in addition to the installation of new hardware on those sites that can be used dually. Only the backhaul and the backbone parts of the network can be used for both existing and new services. Licensees estimate that their cost for building this 3G infrastructure will be at least as high as the license fees. The debt ratio of some of the major carriers, including British

---

structure goes hand in hand with larger quantities supplied.

Telecom, France Telecom, and Deutsche Telekom, has plummeted and resulted in downgrading of their bond ratings, further increasing the cost of network deployment through increased costs of capital.

It can be expected that these factors combined with the collapse of telecommunications and technology stocks will create strong incentives for industry concentration. At least in the short run, the vision of facilities-based competition between 5 or more carriers will have to be abandoned or reconsidered. From our analysis of the GSM and PCS markets we conclude that the license fees will likely create a cost push, especially since they affect all entrants at the same time and with a known magnitude. (In contrast, license fees in the various national GSM markets were more asymmetric as they were imposed at different points in time and not on all carriers.) As our analysis also indicated, a more concentrated market structure does go hand in hand with increased supply (and lower prices). A lot will therefore depend on the net impact of the cost increase due to the license fees and the competitive effect. Likely there is an upper threshold beyond which increased concentration will reduce the intensity of competition and thus work in the same direction as the cost increase from the license fees.

To mitigate this effect and to ease the challenge of financing the infrastructure rollout, the European Union and several national governments have endorsed the concept of network sharing (European Commission 2001; RegTP 2001). A spectrum of options for network sharing is available ranging from the rollout of one joint radio access network to more limited forms of site and antenna sharing. The potential cost savings are highest if one joint network were to be built and lowest in the case of antenna sharing. The emerging consensus seems to approve of sharing of the radio access network as long as the collaborating carrier's networks remain logically separated. According to major suppliers, such sharing could reduce the costs of the radio access network in the initial coverage phase by up to 60% for total savings of up to 40% of the total infrastructure investment cost (Ericsson 2001). Other estimates point to a much lower range for possible savings (Bauer, Westerveld, Maitland 2001). There is a risk that network sharing might facilitate the coordination of pricing, marketing, and other decisions by the major service providers. Moreover, it is unclear whether the claims that networks can be easily unbundled once the market expands are correct.

To avoid negative competitive effects it seems to be advisable to establish clear reporting and monitoring provisions to evaluate the experience with sharing on an ongoing basis and to take mitigating regulatory or antitrust action. The current European framework sets the threshold for adopting ex ante regulatory measures at 25% of the market if there is additional evidence of a dominant market position. The pending discussion on the new framework, to be introduced in 2003 sets this threshold at 50% of the market (European Commission 2000). In the U.S. recent antitrust policy was based on similar criteria although it was not always executed consistently. One area that is commonly neglected is the impact of license fees on the incentive of 3G service providers to enclose access to information in order to increase profits. Such an approach would further undermine the openness of the Internet, already endangered by developments in the cable broadband access market (Bar et. al. 2000).

Despite the fact that license fees impact the subsequent evolution of the market, workable alternatives to spectrum pricing are not yet fully developed. In the short run, the potential negative impact can be reduced by appropriately shaping the legal framework of the 3G industry. For example, upfront license fees could be converted into long-term lease payments thus reducing the concentration incentive (Noam 1998). Some countries, including Denmark, have chosen this approach. Likewise, the establishment of secondary markets for spectrum would reduce the consolidation pressure. In the long run radically new approaches to spectrum management may be called for. Technological developments in the spread spectrum area may facilitate the creation of fully open spectrum markets and the treatment of spectrum as a commons (Benkler 2001).[5] However, currently the transaction costs of such an approach if high-quality service is to be provided are likely very high and possibly prohibitive. Nevertheless, the experience with 3G auctions ought to stimulate the discussion on meaningful alternatives, especially in nations that have not yet determined their approach.

## 6. Conclusions

The evidence from the GSM and PCS markets that is reviewed in this paper suggests that, contrary to a widely held belief, market entry feed do indeed influence the subsequent development of the market. We discuss three potential transmission channels by which license fees can influence the price and quantity of service sold in a wireless market: an increase in average cost, an increase in incremental costs, and impacts of sunk costs on the emerging market structure. From this conceptual debate, an empirical model is developed and tested using cross-sectional data. We utilize a structural equation approach, modeling the supply and demand relationships subject to the constraint that supply equals demand. The results confirm the existence of a positive effect of license fees on the cost of supply. However, we also find that higher market concentration has a positive effect on the overall supply in the market, perhaps supporting a Schumpeterian view that a certain degree of market concentration facilitates efficiency. For the emerging 3G market this implies that the high license fees paid in some European markets will likely increase the trend towards industry consolidation. This will not necessarily result in a worsening of the overall performance of service providers. However, close monitoring of the evolution of the 3G market will be necessary to detect possible abuses of market power early on. In the medium and long run it would be desirable to engage in a debate on alternatives to spectrum auctions.

---

[5] For a critical discussion of open spectrum models see Brennan (1998).

Table 1

**GSM and PCS licensing conditions in selected OECD countries**

| Country | Initial payment ($000) | Recurring annual fees ($000) | Total after 5 years ($000) | Fee per subscriber ($) | Licensing method |
|---|---|---|---|---|---|
| Austria, GSM, (Mobilkom) | 258,117 | 242 | 260,432 | 100.17 | Administrative |
| Belgium, GSM (Belgacom) | 198,882 | 900 | 203,382 | 195.56 | Administrative |
| Denmark, GSM (TeleDanmark) | 0 | 30 | 150 | 0.11 | Administrative |
| France, GSM (France Telecom) | 135 | 5,969 | 29,978 | 2.97 | Administrative |
| Germany, GSM (Deutsche Telekom) | 0 | 0 | 0 | 0 | Administrative |
| Netherlands, GSM (KPN) | 0 | 95 | 475 | 0.14 | Administrative |
| US, Sprint PCS | 2,645,278 | 0 | 2,645,278 | 465.66 | Auction |

Table 2

**3G licensing conditions in selected OECD countries**

(as of September 2001)

| Country | No. of licenses | License fee ($000) | Fee per subscriber ($) | Obligations | Licensing method |
|---|---|---|---|---|---|
| Austria | 6 | 720,000 | 171 | 25% coverage by 2003, 50% by 2005 | SMR auction |
| Denmark | 4 | 472,000 | 180 | 30% coverage by 2004, 80% by 2008 | Sealed bid auction |
| Finland | 4 + 2(1) | No license fee | 0 | No coverage obligations | Beauty contest |
| France | 2 of 4(2) | 9,076,000 | 880 | 80% coverage by 2009 | Beauty contest |
| Germany | 6 | 45,720,000 | 1,948 | 25% coverage by 2003, 50% by 2005 | SMR auction |
| Italy | 5 | 10,980,000 | 365 | regional capitals within 30 months | Auction with beauty contest in pre-qualification stage |
| Netherlands | 5 | 2,240,000 | 330 | 60% coverage by 2008 | SMR auction |
| UK | 5 | 34,200,000 | 1,430 | 80% of population by 2008 | SMR auction |
| US | TBD | N/A | N/A | N/A | Not yet assigned |

(1) Four national licenses plus two regional licenses.
(2) In France due to a lack of bidders only 2 of the 4 licenses were awarded early in 2001. ART, the national regulatory authority plans to open bidding for the two other licenses again in 2002.

Sources: GSM Europe, OECD 2001, own calculations.

Table 3

**Summary statistics**

| Variable | Observations | Mean | Standard Deviation | Minimum | Maximum |
|---|---|---|---|---|---|
| $p^W$ | 18 | 181.72 | 116.46 | 29 | 396 |
| $q^S$ | 18 | 0.225 | 0.115 | 0.02 | 0.45 |
| $C_L$ | 18 | 55.54 | 115.08 | 0 | 465.66 |
| COMP | 18 | 4356.89 | 1302.28 | 604 | 6084 |
| POPD | 18 | 136.56 | 108.03 | 14 | 388 |
| W | 18 | 14.71 | 5.13 | 3.64 | 24 |
| $p^F$ | 18 | 337.88 | 62.44 | 239.69 | 490.65 |
| INC | 18 | 25944.72 | 8496.76 | 11385 | 44962 |
| $TD^F$ | 18 | 58.35 | 11.72 | 42.1 | 76 |

Table 4

**Correlation matrix**

|  | $q^s$ | $p^W$ | $C^L$ | COMP | POPD | W | $p^F$ | INC | $TD^F$ |
|---|---|---|---|---|---|---|---|---|---|
| $q^s$ | 1.0000 | | | | | | | | |
| $p^W$ | -.5119 | 1.0000 | | | | | | | |
| $C_L$ | -.5295 | .2537 | 1.000 | | | | | | |
| COMP | .6246 | -.3649 | -.5103 | 1.0000 | | | | | |
| POPD | -.3476 | .0063 | -.0836 | -.0470 | 1.0000 | | | | |
| W | .4543 | -.4825 | -.0470 | .2276 | .1355 | 1.0000 | | | |
| $p^F$ | -.2232 | .6606 | .0703 | -.0383 | -.0489 | -.6059 | 1.0000 | | |
| INC | .2687 | -.4366 | .2224 | -.0087 | -.0050 | .7618 | -.8000 | 1.0000 | |
| $TD^F$ | .1251 | -.5444 | .1324 | -.1357 | -.0551 | .5907 | -.8079 | .8179 | 1.0000 |

Table 5

**3SLS structural equation estimation results**

|  | Coefficient | Standard error | z | P > \|z\| |
|---|---|---|---|---|
| **Supply** | | | | |
| $p^W$ | .0003566 | .0003044 | 1.172 | 0.241 |
| $C_L$ | -.0004917 | .0001672 | -2.940 | 0.003 |
| COMP | .0000299 | .0000156 | 1.922 | 0.055 |
| POPD | -.0005243 | .000152 | -3.449 | 0.001 |
| W | .0126294 | .0041019 | 3.079 | 0.002 |
| Constant | -.0855854 | .1602322 | -0.534 | 0.593 |
| $R^2$  0.6198 | | $P(\chi^2)$  0.0000 | | |
| **Demand** | | | | |
| $p^W$ | -.001698 | .0006385 | -2.659 | 0.008 |
| INC | .000013 | 6.02e-06 | 2.161 | 0.031 |
| $p^F$ | .0016634 | .0010406 | 1.599 | 0.110 |
| $TD^F$ | -.0066927 | .0041347 | -1.619 | 0.106 |
| Constant | .2181322 | .4305321 | 0.507 | 0.612 |
| $R^2$  0.2156 | | $P(\chi^2)$  0.0432 | | |

Figure 1

**Impact of license fees on average costs and market outcomes**

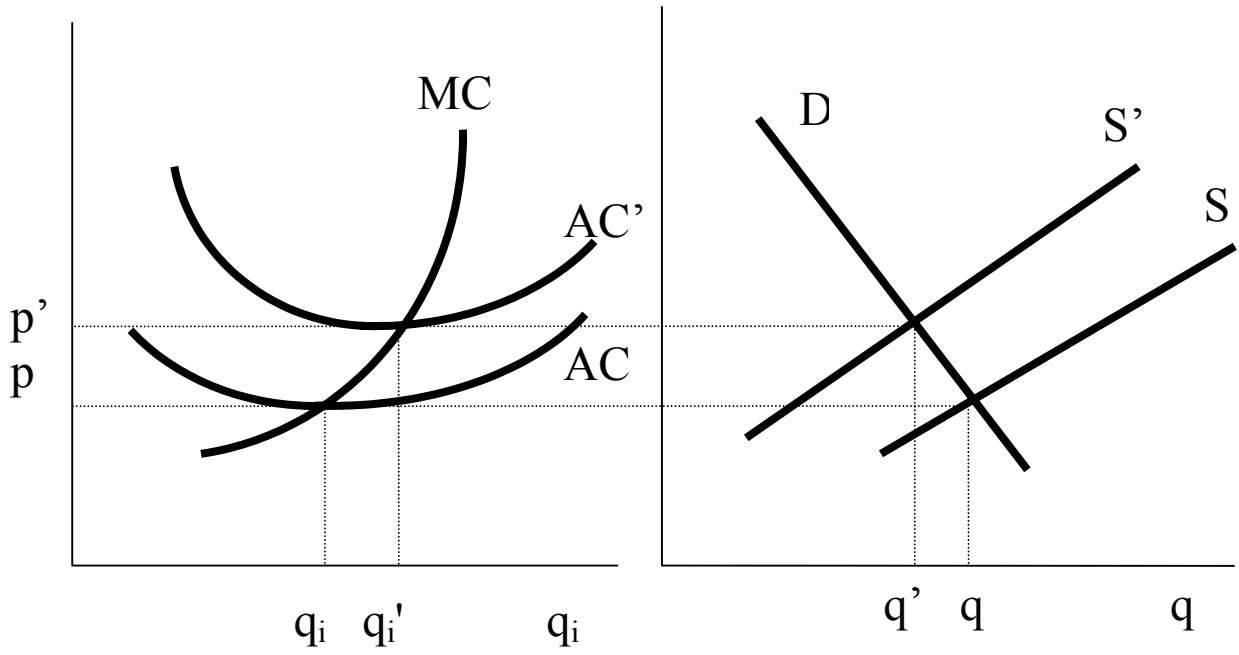

Figure 2

**Impact of variations in variable cost**

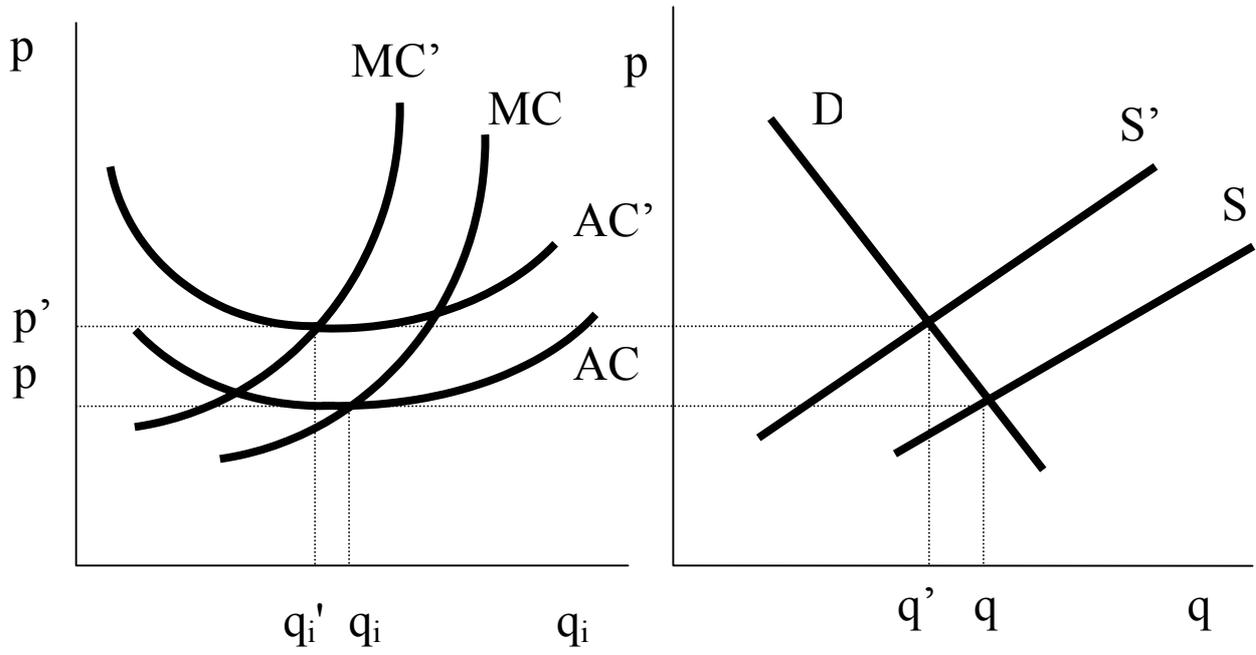